\documentstyle[epsfig]{aipproc}

\newbox\grsign \setbox\grsign=\hbox{$>$} \newdimen\grdimen \grdimen=\ht\grsign 
\newbox\simlessbox \newbox\simgreatbox \newbox\simpropbox
\setbox\simgreatbox=\hbox{\raise.5ex\hbox{$>$}\llap
     {\lower.5ex\hbox{$\sim$}}}\ht1=\grdimen\dp1=0pt
\setbox\simlessbox=\hbox{\raise.5ex\hbox{$<$}\llap
     {\lower.5ex\hbox{$\sim$}}}\ht2=\grdimen\dp2=0pt
\setbox\simpropbox=\hbox{\raise.5ex\hbox{$\propto$}\llap
     {\lower.5ex\hbox{$\sim$}}}\ht2=\grdimen\dp2=0pt
\def\simgreat{\mathrel{\copy\simgreatbox}}
\def\simless{\mathrel{\copy\simlessbox}}

\newcommand{\bc}{\begin{center}}
\newcommand{\ec}{\end{center}}
\newcommand{\be}{\begin{equation}}
\newcommand{\ee}{\end{equation}}
\newcommand{\beq}{\begin{eqnarray}}
\newcommand{\eeq}{\end{eqnarray}}
\newcommand{\bez}{\begin{eqnarray*}}
\newcommand{\eez}{\end{eqnarray*}}

\def\etal{\mbox{\it et al.}}

\def\Tbb{T_{\rm bb}}
\def\kTe{kT_e}

\def\ergs{\;{\rm ergs}}
\def\secinv{\;{\rm s}^{-1}}

\begin{document}

\title{Modeling X/$\gamma$-ray Spectra of Galactic Black Holes and Seyferts}

\author{Juri Poutanen} 
\address{Stockholm  Observatory, SE-13336 \ Saltsj\"obaden, Sweden} 

\maketitle

\begin{abstract}
Recent high quality broad-band X-ray and gamma-ray observations of Galactic black 
holes by {\em ASCA, Ginga, RXTE}, and {\em CGRO} show that the spectra cannot be 
described by simple thermal Comptonization models (even including a fully 
relativistic treatment). Multi-component models 
accounting for Comptonization by thermal and non-thermal electrons, 
Compton reflection of X/$\gamma$-rays from cold matter, and 
emission of the optically thick accretion disk are required.
In this review, I discuss recent advances in the spectral modeling of Galactic 
black holes and Seyferts and the codes available for such an analysis.
\end{abstract}

\section*{Spectral models} 

Already at the time of the discoveries in the early 1970s of the first X-rays 
sources it was realized that the most efficient process to produce 
large X-ray/$\gamma$-ray (X$\gamma$ hereafter) luminosities is Comptonization
of soft seed photons by a hot rarefied electron gas (e.g., \cite{sle76}). 
Detailed calculations of thermal Comptonization spectra were 
carried out by Monte-Carlo \cite{gw84,pss83,stern95a} and analytical methods 
\cite{sle76,is72,fr72,st80}. Analytical formulae describing spectra emerging 
from an optically thick 
cloud of non-relativistic electrons have been derived by solving 
the Kompaneets equation \cite{komp57} in a ``double'' diffusion approximation 
(in optical depth and frequency) \cite{st80} and 
have been used extensively during last twenty years to fit the X$\gamma$ data. 
Only recently the data has become good enough to require a more accurate theoretical
description of Comptonized spectra. The usefulness of generalizations of 
the analytical
formulae to a mildly-relativistic gas and an optically thin media \cite{tit94} 
is disputable \cite{zjm96} and not trivial, since both electron 
and photon energies can be of order $\sim m_ec^2$.  
No analytical formulae exist for Comptonized  spectra for 
non-Maxwellian electron distribution (except for  power-law 
distributions of electrons in the optically thin limit). 
Monte-Carlo simulations
unfortunately, are still too slow to be implemented
into the standard software for the X$\gamma$ data analysis (e.g. XSPEC), 
but on the other 
hand give the possibility to consider arbitrary geometries.  

All this makes attractive the use of numerical methods capable in 
solving relativistic kinetic equations for Compton scattering on 
electrons of arbitrary energy distribution 
without any restrictions on photon and electron energies 
in a reasonably short computer time. Such codes have become available 
recently \cite{coppi92,ps96}. Some codes (XSPEC versions 
{\sc nteea} \cite{zdz90} and {\sc eqpair} \cite{czm98}) 
follow not only the photon, but also the 
electron distribution. They also allow, in a self-consistent manner,  
the determination of the electron temperature, optical depth, and 
such physical quantities as, e.g., the heating rate of thermal
particles, the efficiency of acceleration of nonthermal particles, 
directly from observations. 

An important ingredient of spectral modeling is an accurate treatment 
of the reflection of X$\gamma$-rays from the cold material (``Compton 
reflection'')  in the vicinity of the compact object  (accretion disk) 
\cite{wlz88}. 
The reflected spectrum depends on a number of factors: 
ionization state of the material, element abundances, 
angular distribution of the incident photons, and, of course, their spectral 
distribution (and polarization).  
A number of models simulating Compton reflection is implemented 
into XSPEC. The most developed model is probably {\sc pexriv} \cite{mz95}. It 
incorporates exact Green's functions accounting for the angular 
dependence of the outgoing (observed) reflected spectrum and different ionization
states of the reflecting medium, while being self-consistent 
regarding the ionization balance 
only up to a reflector temperature of order 0.1 keV \cite{done92}. 
If the incident spectrum is anisotropic as expected, e.g.,  
in the case of magnetic flares on the surface of accretion disks 
\cite{hmg94,stern95b,sve96a,sve96b}, the reflected spectrum should be computed
using Green's functions that account for angular dependence of both incident and 
reflected radiation \cite{pns96}.

Rotation of the accretion disk can also be an important factor 
causing measurable distortion of the reflected spectrum. An exact solution 
for a Kerr black hole is far from being simple and is not yet available 
in XSPEC. 
A useful approximation, the convolution of the 
Compton reflection from the static matter with a relativistic disk
line profile \cite{fab89}, has already been used to interpret smeared
iron edges in some Galactic black holes (GBH) \cite{gier97a,zycki97a}. 

\section*{Seyferts and  the hard state of GBH} 

Historically, the thermal Comptonization model was used to fit 
the data from GBH (see, e.g., \cite{st80,st79,g93}).
While the spectra of Seyfert galaxies were explained in terms of the 
nonthermal pair-reflection model \cite{sve94}, which well reproduced 
a quite small spread in the observed X-ray spectral indices \cite{np94}. 
The fact that the X-ray spectra of GBH in their hard state are quite similar 
to the spectra of Seyferts was realized quite early \cite{wfm84}, while 
only recently they were shown to be similar also up to the soft $\gamma$-rays 
\cite{zdzint}. Probably, 
the only difference is that amount of Compton reflection is generally higher 
in Seyferts than in GBH \cite{gon96,ebis96,gier97b}. 
Both GBH and Seyferts show a cutoff in their spectra at $\sim 100$ keV, thus
making thermal models more likely. 
The best available data yield a Thomson optical depth, 
$\tau\approx 1$, and a temperature, $\kTe\approx 50-100$ keV, of the Comptonizing 
cloud. 

\begin{figure} 
\centerline{\epsfig{file=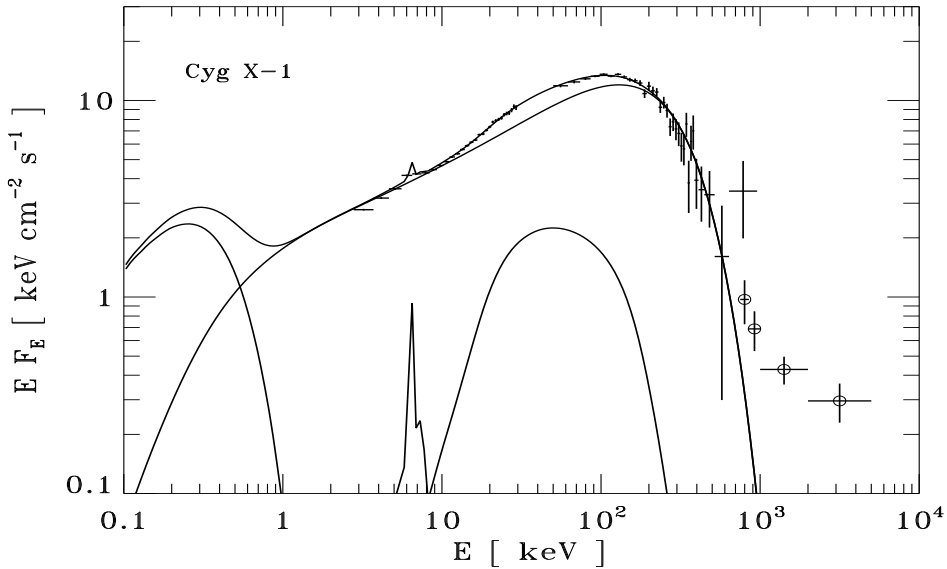,width=5.2in,height=3.in}}
\vspace{10pt}
\caption{The hard state of Cygnus X-1. Simultaneous data from
{\em Ginga}, OSSE and COMPTEL in June 1991. High energy excess
is clearly visible in the COMPTEL data at $E\simgreat$ 1 MeV.
Different components correspond to the soft seed radiation from the accretion 
disk (or cold clouds mixed with the hot plasma), thermal Comptonized spectrum, and
its Compton reflection (together with the fluorescent iron line) 
from a weakly ionized cold matter. 
}\label{fig:cygx1hs}
\end{figure}

Evidence for deviations from the (intrinsic) pure thermal Comptonization spectrum 
is rather weak in the case of Seyferts. None of them has been detected 
above $\sim 500$ keV with modern instruments \cite{mai95,joh97}.  
The fraction of power that can be channeled to the 
nonthermal particles (nonthermal efficiency) is $\simless 10$ per cent in 
the case of NGC 4151, the brightest Seyfert galaxy in X$\gamma$-rays 
\cite{zjm96}.
On the other hand, observations by BATSE and COMPTEL revealed the presence
of a high energy excess at $\simgreat$ 500 keV in two GBH,  Cygnus X-1 and 
GRO J0422+32 \cite{mcc94,ling97,dijk95} (see Fig.~\ref{fig:cygx1hs}), that 
requires $\sim$ 10 per cent nonthermal efficiency \cite{lkl96,pc98}. 
Electrons have an almost Maxwellian distribution with
a weak power-law tail.
Deviations from a Maxwellian distribution could also 
explain why a combination of two thermal Comptonization spectra was 
required to fit the data of Cyg X-1 \cite{gier97b}, while a single 
component corresponding to Comptonization by a hybrid plasma 
\cite{coppi92} provides a very good fit to the 
data from 1 keV up to 4 MeV (see \cite{pc98} and 
Fig.~\ref{fig:hybrid}). 
 
What is the geometrical arrangement of the hot and the cold matter? 
In a simple slab-corona model \cite{liang79,hm93}, hot corona covers the whole 
cold disk. When most of the energy is dissipated in the hot phase, 
soft seed photons for Comptonization are produced only by reprocessing X$\gamma$ 
radiation and the resulting spectrum is the hardest possible. However, 
it was recently shown that even these spectra are too steep to be consistent with 
observations of GBH and Seyferts \cite{stern95b,pkr97,dove97,zdz98b}. 
In principle, the spectra from the slab-corona can be rather flat in 
the standard X-ray region due to the anisotropy effects \cite{ps96,hm93} 
which become important for high electron temperature, 
but then the high energy cutoffs are not compatible with the data. 
Additional soft photons (besides the reprocessed ones), 
that could be generated 
in the cold disk, worsen the discrepancy. 

Localized active regions (magnetic flares) on the surface of the accretion 
disk \cite{hmg94,gal79} have  less soft 
photons returning to the region than the hot slab has. 
Thus, being more photon starved, they produce
harder spectra in consistence with observation. 
Due to the anisotropy of the radiation scattered once in the corona, 
the spectrum should have a break ({\em anisotropy break}, see 
\cite{stern95b,sve96a,sve96b}), which has never been observed. 
In the case of Seyferts, which can have smaller temperature of the seed photons, 
this break can be hidden below $\sim$ 1 keV making its detection impossible 
due to interstellar absorption. The amount of Compton reflection in GBH
is much smaller than expected from flares atop of an  extended cold 
disk making this model questionable, at least for GBH (however, see  
\cite{nm97}). 

A hot inner disk and a cold outer disk model (e.g. \cite{sle76,bkb77}) 
would be compatible with both the observed spectral indices and 
the amount of Compton reflection.
If the cold disk provides soft photons for cooling then, in the case of 
Cyg X-1 and GX339-4,  its inner edge 
should penetrate into the hot region \cite{pkr97,zdz98b} in order to provide 
enough reprocessed soft photons. 
The data are also compatible with a model where 
small dense cold clouds mixed with the hot media 
\cite{light74,cel92,col96,kun97} reprocess X$\gamma$ radiation and provide
seed soft photons \cite{zdz98b}. In any case, the outer disk should be 
flared in order to give rise  to the observed amount of Compton reflection. 
As was already mentioned, Seyferts generally 
have a larger observed amplitude of Compton reflection than GBH. This can be 
related either to the different structure of the inner accretion disk, or 
just to the fact that a large amount of cool material (the molecular torus) 
\cite{kmz94,ghm94} surrounds the accretion disk in Seyfert galaxies 
providing additional reflection. Concluding, these models can unify the 
geometry of GBH and Seyferts' accretion disks.

It is quite difficult to quantify spectral variability even of the brightest GBH in 
the soft $\gamma$-rays, since the minimum integration time (with OSSE) needed to get 
a spectrum with a reasonable statistic at $\simgreat 100$ keV is about one hour.  
In the hard state, there are evidences that Cyg X-1 shows an increase of the cutoff 
energy when the luminosity in the hard X-rays 
drops, while the spectrum in the standard X-ray band, 2--20 keV, hardly changes
\cite{gier97b}. Changes in the optical depth, $\tau$, 
and electron temperature compensate each other in a way to keep 
the Kompaneets $y$-parameter constant. Since the 
temperature increases when the luminosity decreases, 
the spectra for different luminosities 
cross each other at about $\sim 500$ keV. 
Such a behavior implies a constant ratio of the seed soft luminosity 
to the heating rate of the hot cloud. This is possible 
if the transition radius between hot and cold disks does not change much 
in the case the seed photons are internally  produced in the outer cold disk.
Alternatively, the  seed photons could be provided mostly by reprocessing of the 
X/$\gamma$-rays.

GBH are highly variable on all possible 
time scales (see, e.g., \cite{klis95}). Of course, a successful 
model must explain both spectral and temporal properties. 
The size of the X$\gamma$ source inferred from the time lags between 
hard and soft photons \cite{miy91,cui97} is a factor of $\sim 100$ larger than 
the size inferred from spectral modeling, assuming that the lags 
are related to the scattering time. 
One possible solution is to propose  a 
large ($\sim 10^4$ gravitational radii) size of the Comptonizing cloud 
\cite{kaz97,hua97}, but it is difficult to imagine any mechanism that could 
heat the plasma up to 50--100 keV and 
dissipate most of the energy at such large distances from
the black hole. On the other hand, it is possible that the time lags
are related not to the scattering time, but, e.g., to the life-time of the 
magnetic flares that may be responsible for the energy dissipation.

\section*{The soft state of GBH} 

\begin{figure} 
\centerline{\epsfig{file=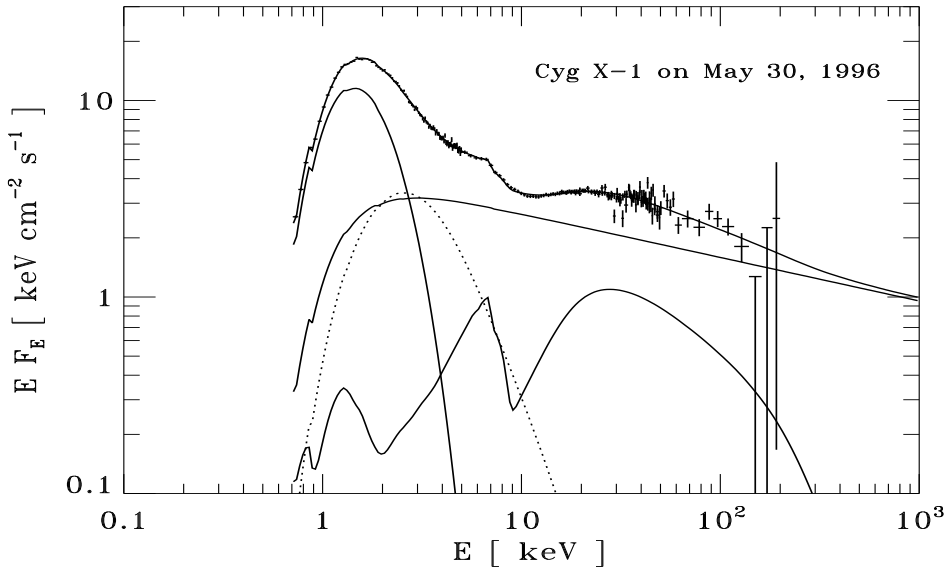,width=5.2in,height=3.in}} 
\vspace{10pt}
\caption{The soft state of Cygnus X-1. Simultaneous observations by
{\em ASCA} and {\em RXTE} on May 30, 1996 \protect\cite{gier98}. 
The overall spectrum consists of disk emission, a non-thermal
power-law, Compton reflection from an ionized disk, and an additional
(thermal Comptonization) component peaking at $\sim$ 3 keV.
}\label{fig:cygx1may96}
\end{figure}

Unlike in the hard state, most of the luminosity in the soft state is carried
by a blackbody-like component with a characteristic temperature,
$k\Tbb\approx 0.3-1$ keV. 
Figure~\ref{fig:cygx1may96} gives an example of the soft state spectrum of Cyg X-1
observed by {\em ASCA} and {\em RXTE} in 1996.
A soft component (at $\simless 5$ keV)
cannot be represented by a black body, or by a 
multicolor disk spectrum, or by a modified black body. It is clear
that at least two component are required to fit it (e.g., a  black body and
a power-law, or two black bodies \cite{gier97a,cui97a}). 
The soft black body probably originates in the accretion disk.
The nature of the additional component peaking at $\sim 3$ keV is not
so clear. Gierli\'nski \etal\/ \cite{gier97a} 
interpreted it as due to the thermal Comptonization
of a disk black body in a plasma with $\kTe\approx 5$ keV and $\tau\approx 3$.

The spectra of GBH in the hard X-rays/soft $\gamma$-rays can be well represented by
a power-law without an observable break at least until $\sim m_ec^2$ 
\cite{grove97a,grove97b,grove98,phlips96}. 
COMPTEL has detected Cyg X-1 and GRO J1655-40
at energies up to $\sim 10$ MeV \cite{iyudin98} 
and it seems that the power-law at MeV energies is just a continuation
of the hard X-ray power-law. If true, these observations 
would rule out the interpretation of the power-law as being due to bulk motion 
Comptonization in a converging flow \cite{etc96,tit97}. This power-law suppose to 
have a break at $\simless m_ec^2$. 
Although the signatures of Compton reflection are also observed
in this state (e.g., \cite{tanaka91}), its magnitude, $R\equiv\Omega/2\pi$,  
is much more difficult to determine since
it depends on the assumed run of ionization with radius and the 
detailed modeling of the continuum which is very curved around the iron edge.
In case of Cyg X-1, $R\approx 0.6-0.8$ \cite{gier97b}, and 
matter providing Compton reflection appears to be strongly ionized 
and the iron edge is smeared by relativistic effects close to 
the black hole. 

The power-law can be produced by Comptonization of soft
photons from the accretion disk by an optically thin nonthermal (probably,
$e^{\pm}$ pair dominated) corona 
which covers much of the disk \cite{pc98,li96a,liang97}. 
In terms of the hybrid thermal/nonthermal model of Coppi \etal\/ 
\cite{coppi92,czm98}, 
the nonthermal efficiency in the soft state is much higher than
in the hard state, while the soft luminosity (from the accretion disk) 
exceeds both the thermal and nonthermal
energy injection to the electrons \cite{pc98}. 
The self-consistent electron distribution can be represented 
by a thermal distribution with a lower temperature (relative to the hard state)
of $\sim$ 30 keV and a significant power-law tail.
The resulting photon spectrum is a black body from
the cold accretion disk with a ``soft excess'' due to
Comptonization by the thermal population of electrons (pairs),  and
a high energy power-law (Comptonization by non-thermal electrons)
extending up to $\sim kT_{bb}\gamma_{max}^2$
(here $\gamma_{max}$ is the maximum electron energy).

\section*{Spectral transitions in Cyg X-1} 

\begin{figure} 
\centerline{\epsfig{file=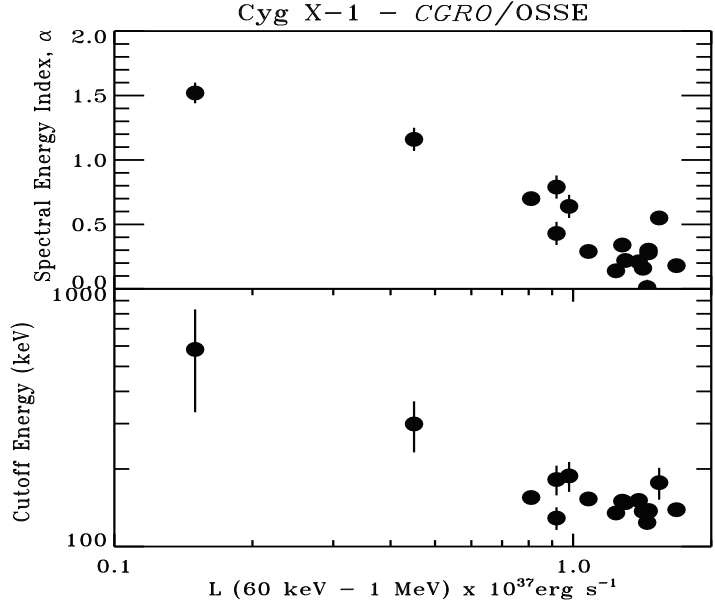,width=5.2in,height=3.in}}
\vspace{10pt}
\caption{
Dependence of the energy spectral index, $\alpha$, 
and cutoff energy, $E_c$, of the best fit model $f(E)\propto 
E^{-\alpha} \exp(-E/E_c)$ 
on the hard X-ray/soft $\gamma$-ray luminosity.  
Data are adapted from Table~3 of Phlips \etal\/ \protect\cite{phlips96}. 
}\label{fig:phlips} 
\end{figure}

What is really happening when a source changes from the hard to the soft state?
In the case of Cyg X-1, when the X$\gamma$ luminosity (1 keV -- 1 MeV) drops below
$\sim 2.5\cdot 10^{37}\ergs\secinv$, 
spectrum in the OSSE range becomes softer and the 
cutoff energy increases (see \cite{phlips96,pacie97} and Fig.~\ref{fig:phlips}).
The hard state OSSE spectra can be well described by a thermal bremsstrahlung 
spectrum even this model is clearly unphysical \cite{sle76}, and 
the bremsstrahlung temperature, $T_{br}$,  has no relation whatsoever to the 
physical electron temperature. 
When the source is in the hard state, $T_{br}$ does not vary much 
\cite{phlips96,kuzn97}. 
During the transition, best fitted $T_{br}$ decreases, but it is difficult 
to interpret this behavior, since thermal bremsstrahlung gives 
very poor fit to the data in the soft state, while a simple power-law 
with an exponential cutoff does a good job giving a superior fit. 

\begin{figure} 
\centerline{\epsfig{file=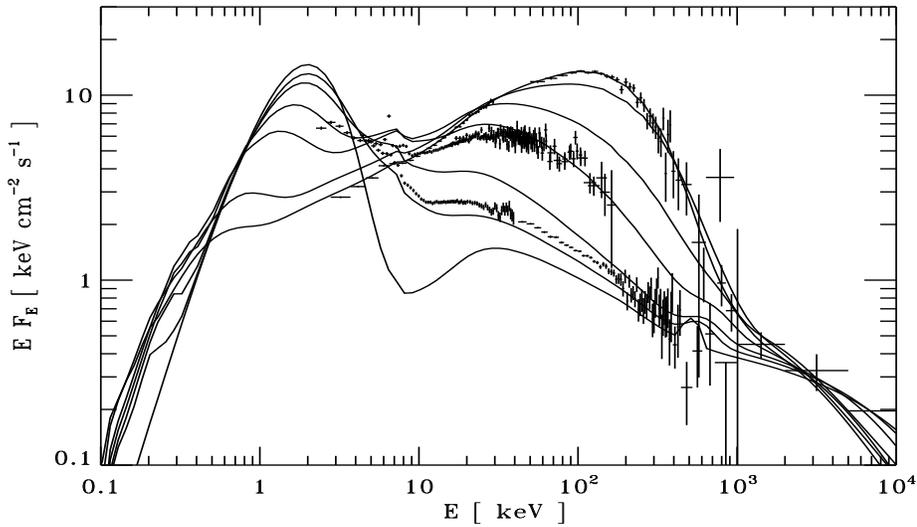,width=5.2in,height=3.in}}
\vspace{10pt}  
\caption{Spectral states of Cyg X-1 and the state transition as
predicted by the hybrid pair model. The sum of cold disk luminosity and 
the thermal dissipation rate in the hot disk remain constant. 
The nonthermal injection rate also does not change during the whole 
transition. Adapted from \protect\cite{pc98,gier98}. 
}\label{fig:hybrid}
\end{figure}

In the hard state,
the inner radius of the cold disk is far away from the black hole
(at $\sim 30-50 GM/c^2$). Almost all the energy is dissipated thermally in
the hot central cloud/corona.
The X/$\gamma$-ray source is photon-starved, 
since the luminosity of the outer cooler part of the accretion flow
is small and the covering factor of the hot corona is quite small too. 
The spectrum is hard. The nonthermal energy injection rate is small as compared 
to the thermal dissipation rate. 
In the soft state, 
the optically thick cool disk moves inwards and receives a 
major part of the dissipated energy \cite{pkr97,min95,esin97}. 
The thermal energy dissipation rate in the
corona is now a factor of $\sim$ 10 lower than the soft luminosity (from
the accretion disk) entering the corona. The rate of energy injection 
to the nonthermal electrons can stay constant during the transition. 
The sum of the soft luminosity from the cold disk and 
the thermal dissipation rate in the hot disk (corona) also remain 
approximately constant. Thus, the whole transition is consistent with 
just a redistribution of the energy input between the 
cold outer disk and the hot inner disk, with a constant nonthermal energy 
injection to the corona (see Fig.~\ref{fig:hybrid}). 

\section*{Summary}

During last twenty years, thermal Comptonization (nonrelativistic) models 
were used extensively to describe the data from accreting black holes. 
Only recently, the observational data became good enough to show that 
these models are only a zeroth order approximation and should be revised 
significantly. We learned that analytical nonrelativistic models 
do not describe the data in the hard X-rays. 
There are growing evidences that the energy distribution of electrons,  
responsible for Comptonizing soft photons, can  deviate notably from a 
Maxwellian. 
Numerical methods 
have appeared that use an exact (fully relativistic) redistribution function 
for Compton scattering and account for accurate radiative transfer and 
the geometry of the Comptonizing cloud.
These methods provide the emission spectrum from such plasma 
as well as a self-consistent electron (-positron) distribution. 

Compton reflection was shown to modify the spectrum significantly and 
smearing of this component due to relativistic effects was observed. 
However, there are still no models for Compton reflection
from a {\em rotating} accretion disk around a {\em rotating} black hole that 
would be implemented into the standard data analysis packages. 

Spectral models considered here are, of course, oversimplified, since
most of them do not consider mechanisms of energy dissipation. 
On the other hand, even the best developed accretion disk models 
still lack important ingredients (e.g. nonthermal power injection) 
that would allow them to describe broad-band data from
the accreting black holes. 
We still lack models that can explain spectral 
and temporal properties together. 

Most importantly, we need more broad-band simultaneous data 
to make firm conclusions about the physics of the accretion disks around 
black holes. 

\section*{Acknowledgments} 

I would like to thank the Swedish Natural Science Research Council and  
the Anna-Greta and Holger Crafoord's Fund for financial support. 
I am grateful to Paolo Coppi, Andrzej Zdziarski, Marek Gierli\'nski, and 
Roland Svensson  
for valuable  discussions and for providing me with some of the data used in this 
review. I thank Roland Svensson  and Felix Ryde for comments on the manuscript.

\end{document}